\documentclass[letterpaper, 10 pt, conference]{ieeeconf}  % Comment this line out if you need a4paper

\IEEEoverridecommandlockouts                              % This command is only needed if 
\usepackage{graphicx} % for pdf, bitmapped graphics files
\usepackage{epsfig} % for postscript graphics files
\usepackage{times} % assumes new font selection scheme installed
\usepackage{amsmath} % assumes amsmath package installed
\usepackage{amssymb}  % assumes amsmath package installed
\usepackage{verbatim}
\usepackage{url}
\title{\LARGE \bf Can a Humanoid Robot be part of the Organizational Workforce?\\A User Study Leveraging Sentiment Analysis}
\author{Nidhi Mishra$^{1}$, Manoj Ramanathan$^{2}$, Ranjan Satapathy$^{3}$, Erik Cambria$^{4}$ and  Nadia Magnenat-Thalmann$^{5}$% <-this % stops a space
\thanks{$^{1}$ Nidhi Mishra is with Institute for Media Innovation, Nanyang Technological University, 50 Nanyang Drive, Singapore 
        {\tt\small nidhi.mishra@ntu.edu.sg}}%
\thanks{$^{2}$ Manoj Ramanathan is with Institute for Media Innovation, Nanyang Technological University, 50 Nanyang Drive, Singapore 
        {\tt\small mramanathan@ntu.edu.sg}}%
\thanks{$^{3}$ Ranjan Satapathy is with Institute for Media Innovation and School of Computer Science and Engineering, Nanyang Technological University, 50 Nanyang Drive, Singapore
        {\tt\small satapathy.ranjan@ntu.edu.sg}}%
\thanks{$^{4}$ Erik Cambria is with  School of Computer Science and Engineering, Nanyang Technological University, 50 Nanyang Drive, Singapore
        {\tt\small cambria@ntu.edu.sg }}%
\thanks{$^{5}$ Nadia Magnenat-Thalmann is with Institute for Media Innovation, Nanyang Technological University, 50 Nanyang Drive, Singapore and MIRALab, University of Geneva, Switzerland
        {\tt\small nadiathalmann@ntu.edu.sg}}%
}

\begin{document}
\maketitle
%\thispagestyle{empty}
%\pagestyle{empty}

%%%%%%%%%%%%%%%%%%%%%%%%%%%%%%%%%%%%%%%%%%%%%%%%%%%%%%%%%%%%%%%%%%%%%%%%%%%%%%%%
\begin{abstract}
Hiring robots for the workplaces is a challenging task as robots have to cater to customer demands, follow organizational protocols and behave with social etiquette. In this study, we propose to have a humanoid social robot, Nadine, as a customer service agent in an open social work environment. The objective of this study is to analyze the effects of humanoid robots on customers at work environment, and see if it can handle social scenarios. We propose to evaluate these objectives through two modes, namely, survey questionnaire and customer feedback. We also propose a novel approach to analyze customer feedback data (text) using sentic computing methods. Specifically, we employ aspect extraction and sentiment analysis to analyze the data. From our framework, we detect sentiment associated to the aspects that mainly concerned the customers during their interaction. This allows us to understand customers expectations and current limitations of robots as employees.

%We placed a social humanoid robot Nadine at an insurance company to evaluate it's capability to work in an open environment and see its performance and acceptance by customers who interacted with her. To evaluate Nadine's capabilities we conducted two different types of studies: In first study we created questionnaire and in second study we asked customers to provide feedback where it was not restricted to any question. We studied both and we show the application of Aspect based sentiment analysis to the humanoid robot's customer feedback. The framework helps in detecting the aspects customers are concerned with and can be used to enhance the human-robot interaction.

\end{abstract}
%%%%%%%%%%%%%%%%%%%%%%%%%%%%%%%%%%%%%%%%%%%%%%%%%%%%%%%%%%%%%%%%%%%%%%%%%%%%%%%%
\section{INTRODUCTION} %%%%%%Manoj

The field of robotics has improved drastically in the last decade. From robots that were initially meant to reduce manual labour or automate menial tasks, current robots focus on social aspects such as teachers ~\cite{Han_EducationRobots_HAI_2015} and companions for the elderly~\cite{Josephine_HealthRobots_SR_2015}. The appearance of robots has also evolved over this period. In recent years, robots which are more human-like and are autonomous when interacting with humans are preferred over conventional robots. Due to this, the field of social robotics has gained momentum. In contrast to early task-based robots, design of humanoid social robots involves developing cognition that considers context, work environment and social interaction clues.

With the development of Artificial Intelligence and robotics, there is a universal question ``Can a humanoid social robot be a part of a company's workforce?''. Does it have all the skills and etiquette to function in an open work environment with different tasks successfully? Human-robot interaction studies are usually conducted in controlled settings or pre-defined tasks. Such kind of interaction would not allow us to understand if a robot can adapt to and perform any role in an 
organization. To answer these questions in this paper, we have conducted an initial study with Nadine~\cite{NadineSR_CGI_2019}, a humanoid social robot as a customer service agent in an insurance company. The setup was entirely open for the public, and the customer could ask any questions. As a service agent, Nadine is required to be able to answer customer queries and maintain a courteous, professional behaviour during her interaction. Our objective was to study if a robot could handle real social situations if customers are willing to interact with human-like robots and study effects of humanoid robots in any work environment.

Nadine's success as an insurance agent is based on the quality of customer-agent interaction. To evaluate the human-robot interaction, we have used survey questionnaires and customer feedback. The survey questionnaire was prepared in such a way that the customer can rate the interaction with the robot agent in terms of functionality, behaviour, usefulness etc. Also, the customer can provide feedback about his/her interaction. Even though surveys can be quantified easily, the same is not applicable for feedback. Usually, customer feedback has to be manually read by someone to understand limitations of the robot. In this paper, we have borrowed sentic-computing~\cite{cambria2015sentic} concepts, specifically, aspect-based sentiment analysis and applied it to the collected customer feedback. In contrast to sentiment analysis, that detects sentiment of overall text, aspect-based analyzes each text to identify various aspects and determine the corresponding sentiment for each one. Using the proposed framework, we can examine each aspect customer talks about. This will also help us to quantify customer feedback data, aspects that customers notice in a work environment. Also, limitations and future extensions to humanoid robots in the work environment can be identified.

The rest of the paper is organized as follows: In section II, we provide related work for robots employed in the work environment and their interaction. We also look into some aspect-based sentiment analysis methods for gauging customer satisfaction. In section III, we explain our experimental setup of Nadine at the insurance company. In section IV, we describe the details of our data collection methods. In section V, we provide details of our framework to analyze user comments based on aspect-based sentiment analysis. In section VI, we present experimental results of the analysis of survey and user comments. We also discuss potential limitations and possible future work. We provide conclusions in section VII.

%The big difference between sentiment analysis and aspect-based sentiment analysis is that the former only detects the sentiment of an overall text, while the latter analyzes each text to identify various aspects and determine the corresponding sentiment for each one. There are two types of aspects defined in aspect-based opinion mining: explicit aspects and implicit aspects. Explicit aspects are words in the opinionated document that explicitly denote the opinion target. For instance, in the above example, the opinion targets screen and resolution are explicitly mentioned in the text. In contrast, an implicit aspect is a concept that represents the opinion target of an opinionated document but which is not specified explicitly in the text. One can infer that the sentence, "This camera is sleek and very affordable" implicitly contains a positive opinion of the aspects appearance and price of the entity camera. These same aspects would be explicit in an equivalent sentence: "The appearance of this camera is sleek and its price is very affordable". We have done aspect based sentiment analysis on collected data.  The results will help make social robots more friendly to humans.

\section{Literature Review}%%%%%%%%Nidhi

\subsection{Robots at work places}
Robots have become an integral part of society. In recent times, several researchers and organizations are considering to make robots a part of their workforce. Initial studies of the robot at workplaces were restricted to simple tasks such as greeting at information booth \cite{Actroid}, performing the predefined skill of archery \cite{iCub:2011} and bartender with communication skills \cite{Giuliani:2013}, museum guide \cite{museum_guide:2005}. Robovie \cite{Tutors_robot:2004} was used as a language tutor for a small group of people in elementary school. In all these tasks, the complexity of human-robot interaction was low, and mistakes made by a robot in such scenarios are inconsequential.

%Aibo and Paro \cite{psychological_enrichment:2004} had animal-like appearances, a dog and a seal respectively. These robots provide entertainment or mental care to people through a human-pet style of interaction.Robovie \cite{Tutors_robot:2004} was used to assist with language education in elementary school. Robovie only interacted with a limited group of people; thus, it is not clear how a robot should operate in large-scale environments where a wide variety of people visit. In all these tasks, the complexity of human-robot interaction was low and effect of robot performing the task badly was inconsequential.

Robots have been considered for a more open work environment with serious consequences as well. For instance, health care \cite{Ljungblad:2012}, restaurant waiters \cite{Robotics_Waiter}, space research \cite{Robonaut} and rescue functionalities \cite{MOIRA} have also been considered. Ljungblad et.al \cite{Ljungblad:2012} introduced a simple utility robot in a hospital environment for transporting goods between departments for 13 days and studied effects of it using interviews, questionnaires and observations. Juan Fasola et.al \cite{Robot_Motivate_Exercise_Older} used a socially assistive robot to motivate physical exercise for older adults. The results of the survey questions regarding participant perceptions and feelings toward the robot were very encouraging; the participants rated the robot highly in terms of intelligence and helpfulness attributed a moderately high level of importance to the exercise sessions, and reported their mood throughout the sessions to be normal-to-moderately pleased. In a restaurant setting, \cite{robot-waiter} employed robots that could move, take orders and even talk to customers in a limited fashion. However, these robots were limited in nature as they could not carry heavier items like soup, pour water to customers or properly communicate. The success of robots in each of these applications is measured differently due to the variation of tasks involved in them. In most of the applications considered for robots at workplaces, the tasks involved are simple and social interactions with the human is considerably less. The appearance of the robots are not human-like always depending upon the task they are involved. In contrast, we choose to use a realistic looking humanoid social robot, Nadine for our experiments. Also, we set her up as an insurance agent to interact with customers in open social scenarios and perform tasks defined for an agent in the organization.

\subsection{Customer satisfaction analysis using aspect-based sentiment analysis}
%In the past, this work must be done by collecting customer's feedback's from the investigation questionnaires, which very time consuming and customers usually like to skip them. However, the volume of these reviews is too heavy for human to read reviews one by one. Therefore, having automation to find weakness intelligently automatically becomes exceedingly valuable work. In this paper we use 

Hu et. al \cite{Mining_summarizing_customer_reviews} analyzed the customer reviews using an aspect extraction method. The authors restricted themselves to explicit aspects and set of rules based on statistical observations. Scaffidi et. al \cite{Product_feature_Scoring_from_Reviews} presented a method that uses a language model to identify product features. They assumed that product features are more frequent in product reviews than in a general natural language text. However, their method seems to have low precision since retrieved aspects are affected by noise.

%Aspect extraction from opinions was first studied by Hu and Liu \cite{Mining_summarizing_customer_reviews}. They introduced the distinction between explicit and implicit aspects. However, the authors only dealt with explicit aspects and used a set of rules based on statistical observations.

Zhang et. al \cite{Weakness_Finder} introduced such an expert system, Weakness Finder, which can help manufacturers find their product weakness from Chinese reviews by using aspect-based sentiment analysis. For explicit features, they incorporated a morpheme-based method and Hownet based similarity measure for grouping them. While a collocation selection method for each aspect was employed for grouping implicit features. For each of the extracted aspects, they utilized a sentence based sentiment analysis method to determine the polarity. All these methods were applied to product reviews to quantify about product's features and usage. In contrast, we apply sentic-computing to understand customer demands and expectations of humanoid robot agent in a work place.
%Weakness Finder extracts the features and groups explicit features by using morpheme based method and Hownet based similarity measure, and identify and group the implicit features with collocation selection method for each aspect. Then utilize sentence based sentiment analysis method to determine the polarity of each aspect in sentences. 
%Sentic computing is a state-of-the-art in all the sentiment analysis applications, whose novelty gravitates around multi-disciplinarity, semantics and linguistics rather than mono-disciplinarity, syntax and statistics. Sentic computing deals with multi-word (concepts) expressions~\cite{Rajagopal2013} rather than words.

For any new technology, gauging customer satisfaction is very important as it can provide essential insights on usefulness and customer demand. For these reasons, customers are usually asked to rate their experience via simple questionnaires and provide feedback. However, customers tend to skip these surveys as it is usually voluntary. The analysis of such feedback comments is also tedious as it requires someone to read all reviews and highlight the primary customer demands manually. Such manual analysis could be time-consuming and subject to human bias. In contrast, in this paper, we propose a NLP based framework relying on aspect-based sentiment analysis to analyze customer feedback and to get insights on Nadine's performance as an agent, overall customer experience with her and areas for improvement.

\section{Experimental Setup}
\begin{figure}[h]
    \centering
    \includegraphics[width = 0.4\textwidth]{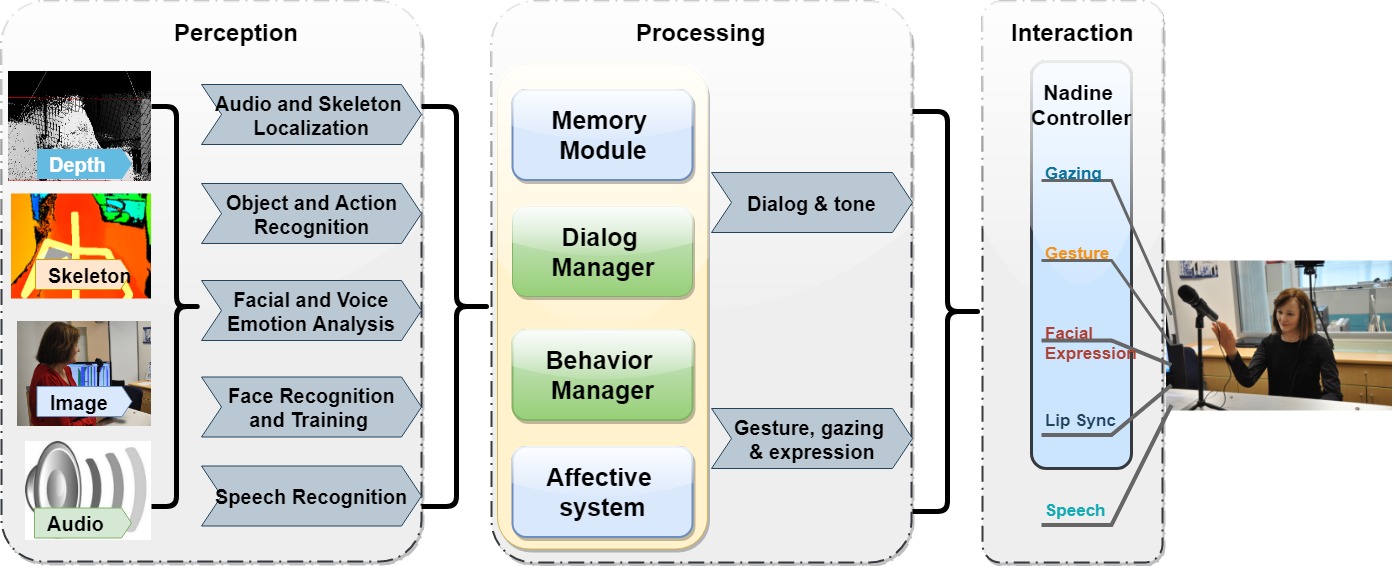}
    \caption{Nadine Robot's Architecture}
    \label{fig:framework}
\end{figure}

%\subsection{Robot Architecture}
For our experiments, we have used Nadine, a realistic humanoid social robot with natural skin, hair and appearance. Figure \ref{fig:framework} shows Nadine's architecture that consists of three layers, namely, perception, processing and interaction. Nadine \cite{NadineSR_CGI_2019} receives audio and visual stimuli from microphone, 3D cameras and web cameras to perceive user characteristics and her environment, which are then sent to the processing layer.
%Perception layer receives the raw stimuli from different modalities and that helps Nadine understand the context and her environment of operation. In Nadine, a Microsoft Kinect, webcam and microphone are used for providing vision and audio inputs in their raw form.  Using these inputs, each sub-module outputs several user characteristics such as identity, gender, emotion, gestures etc and environmental characteristics such as objects. These outputs are given as input to be processed by the next layer (processing). 
The processing layer is the core module of Nadine that receives all results from the perception layer about environment and user to act upon them. This layer includes various sub-modules such as dialog processing (chatbot), affective system (emotions, personality, mood), Nadine's memory of previous encounters with users. Responses are sent to interaction layer so that they can be executed and visibly shown by Nadine such as head movement to maintain eye gaze, gestures and facial expressions, dialog and tone (to show different emotions, personality).

\begin{figure}[h]
    \centering
    \includegraphics[width =0.5\textwidth]{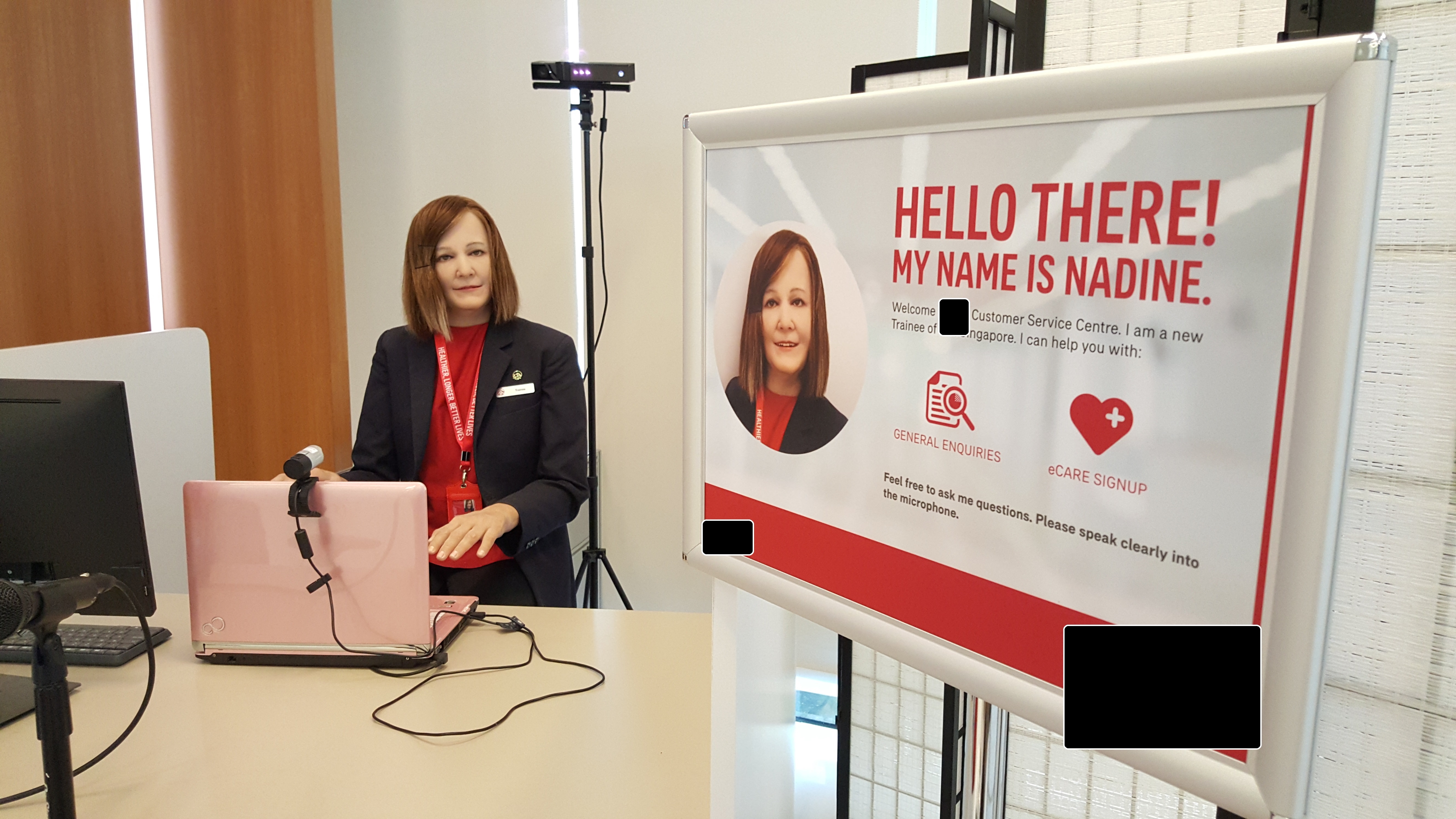}
    \caption{Nadine setup in insurance company}
    \label{fig:Nadine_setup}
\end{figure}

Nadine was set up in an insurance company to work as a customer service agent alongside other human employees. She was required to handle customer queries and behave like a courteous service agent. For customer queries, the company had provided several FAQs from their customer interactions. A separate chatbot \cite{Chatterbot_2018} was trained based on these FAQs and integrated into Nadine, that allowed her to handle customer queries. The main objective of the study was to see if customers were willing to interact with the robot service agent and is Nadine able to handle such workplace scenarios? The customers voluntarily fill in a survey form to rate their experience with Nadine and provide feedback comments.
%\subsection{Robot setup at workplace}

\section{Data Collection for survey}
To analyze Nadine's performance as a customer service agent, we needed to collect feedback from customers. The data collected was used for analyzing customer-agent(robot) interaction and effectiveness of the robot in the workplace. For this purpose, we employed two modes of data collection, namely, Survey Questionnaire and customer feedback. In this section, we outline the details of both these modalities.
%Before we can start any kind of text analysis, we need to gather information. Below, we will outline the two type different data collection used for the survey.

\subsection{Questionnaire}
We created a questionnaire on Survey Monkey with seven questions. Throughout the questionnaire, Nadine was addressed as staff rather than as a robot. The survey was voluntary for the customers to fill in and was set up in a tablet. We collected $14$ customer survey responses on Nadine's performance as a customer service agent. The questions are tabulated in table \ref{table:cust_survey}
\begin{table}[]
\begin{tabular}{|p{8cm}|}
\hline
What is your gender                                                                                       \\
\hline
How old are you?                                                                                          \\
\hline
Does the staff posses the required skills and knowledge about the company's products and services.           \\
\hline
Was the staff friendly and behaving in a courteous manner when dealing with you.                 \\
\hline
Is the staff is professional and has a pleasing and presentable appearance.                            \\
\hline
Was the staff willing to listen and respond to your needs on time.                                   \\
\hline
How would you rate the ease of access and the usefulness of our online e-care with the help of the staff?\\
\hline
\end{tabular}
\caption{Customer Survey Questions}
\label{table:cust_survey}
\end{table}

\subsection{Customer feedback}
We also asked all customers to give unrestricted feedback on Nadine's performance as a customer service agent at the insurance company so that they can express their opinion on Nadine outside the survey questionnaire. Total of $75$ users gave their valuable feedback on Nadine. These comments were analysed using sentic computing framework explained in the section \ref{Sec:NLP_Framework} and discuss the results of the analysis in section \ref{Sec:Results}.

\section{Proposed NLP Framework for analysis of customer feedback}%%%%%%Ranjan
\begin{figure}[h]
    \centering
    \includegraphics[width =0.5\textwidth]{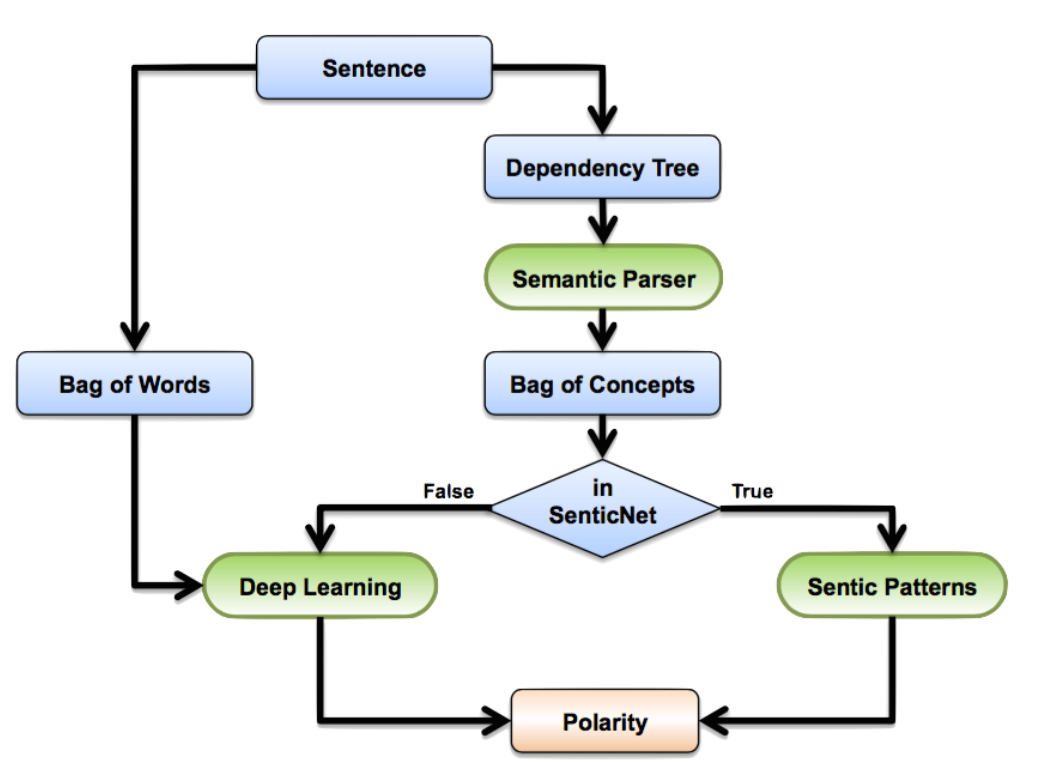}
    \caption{Flowchart of the sentence-level polarity detection framework. Text is first decomposed
into concepts. If these are found in SenticNet, sentic patterns are applied. If none of the concepts
is available in SenticNet, the ELM classifier is employed.~\cite{cambria2015sentic}}
    \label{fig:sentic}
\end{figure}

\label{Sec:NLP_Framework}
This section discusses the proposed framework for the aspect extraction and sentiment analysis on customer comments. We use sentic computing to analyze user comments.  Sentic Computing aims to bridge the gap between statistical NLP and many other disciplines that are necessary for understanding human languages, such as linguistics, commonsense reasoning, and affective computing. The sentic computing framework is designed to receive as input a natural language concept represented according to an M-dimensional space, and predict the corresponding sentic levels for the four affective dimensions. We first analyze the comments based on aspects and then parse the sentence through SenticNet for polarity detection (Figure \ref{fig:sentic}). The framework depicts the sentiment analysis of each aspect customers talk about. As customers talk only about the single aspect in their feedback so, the sentence polarity is the same as the aspect polarity. We explain how aspect extraction and senticnet works in subsection \ref{subsec:aspect} and \ref{subsec:senticnet} respectively.

\subsection{Aspect Extraction}
\label{subsec:aspect}
Aspect-based opinion mining~\cite{hukdd} focuses on the relations between aspects and document polarity. An aspect, also known as an opinion target, is a concept in which the opinion is expressed in the given document.
The framework~\cite{poria2016aspect} incorporates a 7-layer deep convolutional neural network (CNN) which tags each word in opinionated sentences as either aspect or non-aspect word. The model also includes a set of linguistic patterns for the same purpose and combined them with the CNN. This ensemble model is used to extract the aspect from the customers' comments. The procedure for aspect model is as follows:
\begin{enumerate}
    \item form a window around the word to tag
    \item apply CNN on that window
    \item apply maxpool on that window
    \item obtain logits
    \item apply CRF for sequence tagging
\end{enumerate}

The trained model can be found here\footnote{\url{https://github.com/SenticNet/aspect-extraction}}

%\subsection{Sentiment Analysis}
%We analyzed the comments using SenticNet.

%For human annotation we used 3 reviewers, and the procedure and for Deep Learning we used SenticNet, which is a pretrained model on review dataset. Subsection \ref{human} and \ref{senticnet} explains the procedure involved to determine the sentiment an of the comments.
%\subsubsection{Human annotated}
%\label{human}
%We assigned 3 reviewers to tag the comments as positive, negative and neutral. The kappa score we got is 0.38 which is termed fair according to~\cite{fleiss1973equivalence}.

\subsection{SenticNet}
\label{subsec:senticnet}
SenticNet is the knowledge base which the sentic computing~\cite{cambria2015sentic} framework
leverages on for concept-level sentiment analysis. SenticNet is a publicly available semantic resource for concept-level sentiment analysis that exploits an ensemble of graph mining and multi-dimensional scaling
to bridge the conceptual and affective gap between word-level natural language data and the concept-level opinions and sentiments conveyed by them~\cite{cambria2018senticnet}. SenticNet provides the semantics and sentics associated with 100,000 commonsense concepts, instantiated by either single words or multi-word expressions.

%\begin{comment}

%\end{comment}

Figure \ref{fig:sentic} shows how a sentence is processed. The input text is first decomposed into concepts. If these are found in SenticNet~\cite{vilares2018babelsenticnet}, sentic patterns are applied. If none of the concepts is available in SenticNet, the ELM classifier is employed.

%%%%%%%%%%%%%%%%%%%%%%%%%%%%%%%%%%%%%%%%%%%%%%%%%%%%%%%%%%%%%%%%%%%%%%%%%%%%%%%%%%%%%%%%%%%%%%%%%%
\section{Experimental Results and Discussions}
\label{Sec:Results}
In this section, we explain the results of our proposed survey questionnaire and aspect-based sentiment analysis on customer feedback. We also discuss the limitations and possible future directions based on the analysis of the customer-agent interaction data.
%In this section to study the result of survey questionnaire and customer feedback. 

\subsection{Analysis of Questionnaire}
\begin{figure}[ht]
    \centering
    \includegraphics[width =0.5\textwidth]{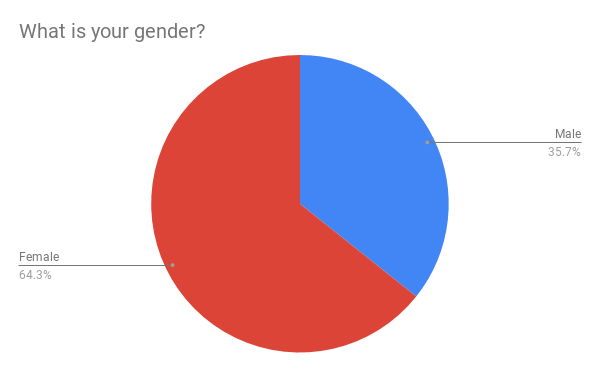}
    \caption{Customer response on question 1}
    \label{fig:q1}
\end{figure}

\begin{figure}[ht]
    \centering
    \includegraphics[width =0.5\textwidth]{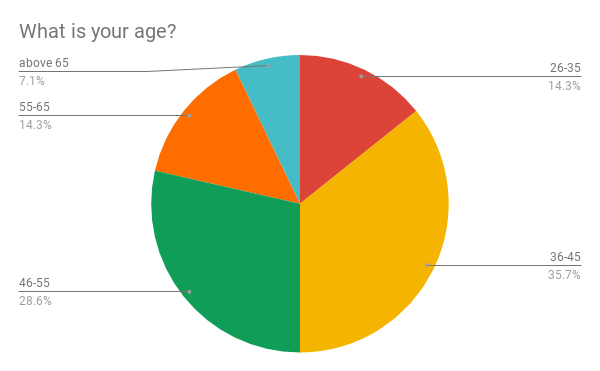}
    \caption{Customer response on question 2}
    \label{fig:q2}
\end{figure}

%The results of questionnaire was computed using 
The first two questions of the survey were meant to understand the customer demographics in the insurance company. This helped us to understand the group of customers that robots like Nadine would attract in a work environment. From our questionnaire, we observed that females were more interested in talking to Nadine. People in the age group of $36 - 45$ had interacted the most with her. In general, we can observe that the younger generation was more comfortable and willing to interact with the robot. The results of both questions can be seen in figures \ref{fig:q1} and  \ref{fig:q2}.

%After analyzing the result of Questionnaire survey, we can see in figure \ref{fig:q1} that females were more interested in talking with our robot Nadine. The age group interacted most was between $36$ to $45$ years is shown in figure \ref{fig:q2}.

It was observed that $39$\% of the customers recorded that Nadine had the required skills and knowledge about the insurance company's products and services. In this initial study, Nadine was required only to answer customer queries and behave with the apt etiquette for the customer service agent. The skill set and questions provided for Nadine were limited. Due to security reasons and privacy concerns, Nadine was not able to access sensitive and personal data such as customer policy information. Due to the open nature of dialog, the customers believed that she could access this type of data and help them with all possible queries, which Nadine was not trained for. This could be the reason why only $32$\% of the customers believed that Nadine was professional. People whose queries she could not handle thought she was not professional. Also, Nadine has a very realistic human-like appearance, which raised customer's expectations of the robot's professional ability and insurance-related functionality to be high. Thus it also shows there needs to trade off between the tasks trained for robots and its appearance.

%As low as $39$ percent people said that she has required skills and knowledge about the insurance company's products and services. The main reason for this could be because Nadine robot did not have access to most of the internal data of Insurance company related to customer because of which she could not help with some of the issues which require more data from insurance company. This was one of the main reason for Nadine to not to be able to help customers completely.

%Only few as $32$ percent customers liked Nadine's appearance, most people did not find Nadine looking professional and presentable. The reason behind this was as Nadine is very human like robot as soon as customer realize that Nadine is a robot, they.........(not sure how to continue)

\begin{figure}[h]
    \centering
    \includegraphics[width =0.5\textwidth]{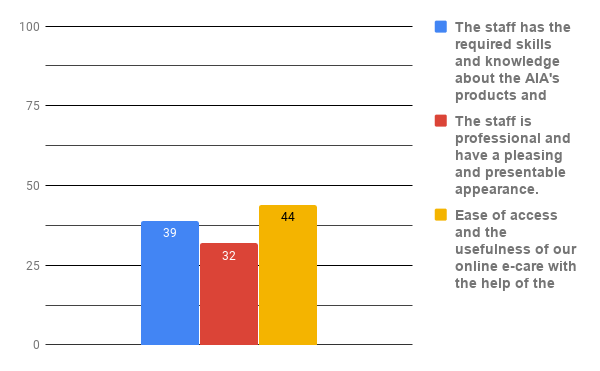}
    \caption{Customer response on questions 3, 5, 7}
    \label{fig:q3}
\end{figure}

Nadine had an additional capability to help customers use the online platform of the insurance company. The motive was to familiarize the customers with the new online platform, which customers can use at their homes to get all routine policy-related information. This would help to avoid unnecessary travel to the service centre. Nadine could guide them step by step to register their account and change their address on online platform. Mostly customers rated Nadine moderate on the help of the online platform. The results of questions 3, 5, 7 are shown in figure \ref{fig:q3}.

%Nadine had an extra capability to help customers use the online platform of insurance company. The motive was to get customers familiar with online platform so that they can solve most of their issues online and next time they do not need to come to the customer service physically. Nadine could guide them step by step to register their account and change their address on online platform. Mostly customers were moderate on help of Nadine on online platform. 

\begin{figure}[h]
    \centering
    \includegraphics[width =0.5\textwidth]{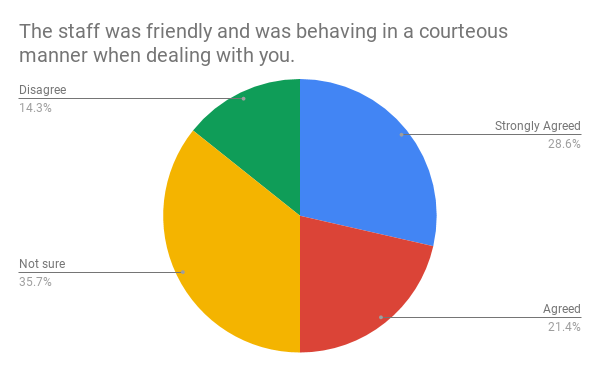}
    \caption{Customer response on question 4}
    \label{fig:q4}
\end{figure}

\begin{figure}[h]
    \centering
    \includegraphics[width =0.5\textwidth]{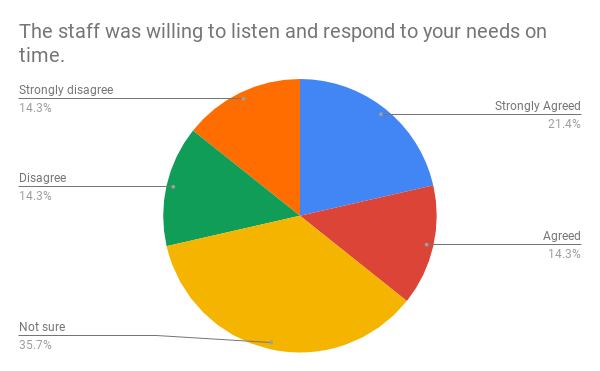}
    \caption{Customer response on question 6}
    \label{fig:q6}
\end{figure}

For the question on Nadine's courteous nature and friendliness, customers mostly agreed or were neutral.  For a customer service agent, it is necessary to be pleasant, courteous, always smiling and show emotions. As a social robot, Nadine has been programmed to simulate a range of emotions and had been set up according to the needs of an agent in the insurance company. Similarly, the customers mostly agreed or were neutral, when asked on Nadine's willingness to listen and respond on time. As a robot, Nadine is always welcoming and willing to listen, but responses could be delayed. For instance, when a customer's question is not in her database, she searches online for the most appropriate answer, which will delay her responses. Also, sometimes, she may not reply when the customer did not speak into the microphone correctly as she did not receive any input. The results of questions 4 and 6 can be seen in figures \ref{fig:q4} and \ref{fig:q6} respectively.

%When asked customers if Nadine was friendly they agreed mostly or they were not sure about it which shows Nadine was able to be courteous. Nadine has an emotion model which lets Nadine to be pleasant in behaviour, smile to customers and show emotion. It is very important for a humanoid robot to have emotion so that robot can feel empathy with customer. The results can be seen in figure \ref{fig:q4}.

%On asking if Nadine was willing to listen and respond on time, the results were mixed. As Nadine being a robot she is always welcoming and willing to listen but sometimes her response to a question can be delayed due to many reason as Nadine may not find an appropriate answer in her own database and to reply now she will search for answer online which can take time. Sometimes she may not reply at all when customer did not speak into microphone which can leave an impression that Nadine is not willing to listen. The results can be seen in figure \ref{fig:q6}.

\subsection{Sentic Computing Framework results of Customer Feedback}
\begin{figure}[h]
    \centering
    \includegraphics[width =0.5\textwidth]{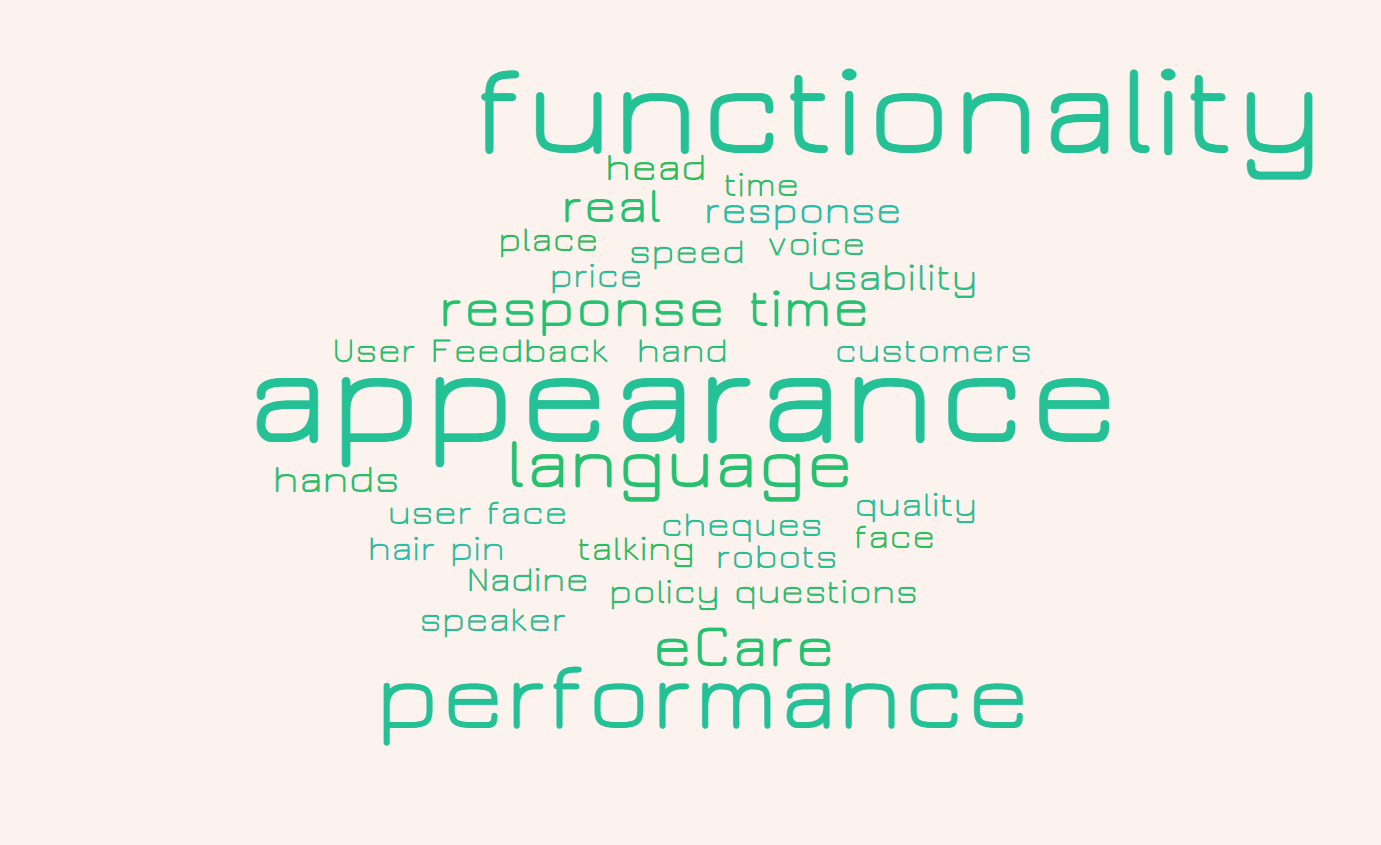}
    \caption{Aspect Cloud from Customer's comments}
    \label{fig:aspect}
\end{figure}

The results show the aspects and sentiments associated with the aspects of the user comments. The results discussed here will help the future generation of humanoid robots to enhance human-robot interaction. The aspects in Figure \ref{fig:aspect} shows the main characteristics that the customer looked for in the robot. In figure \ref{fig:aspect}, the size of the word shows the number of customer feedback about that aspect. It can be observed that functionality, appearance and performance were three main aspects that any customer commented on.

\begin{figure}[h]
    \centering
    \includegraphics[width =0.5\textwidth]{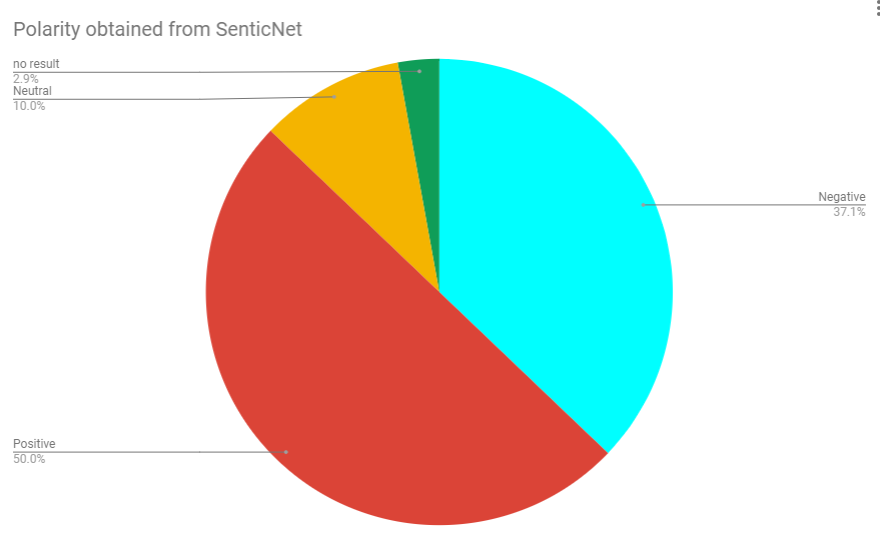}
    \caption{Polarity detection}
    \label{fig:polarity}
\end{figure}

The sentiments can be either Positive, Negative and Neutral. The sentiments observed in customer feedback can be seen in Figure \ref{fig:polarity}. 50\% of the customers were Positive towards Nadine as an employee, 37.1\% Negative and 10\% as Neutral sentiments. SenticNet could not find sentiments in 2.9\% of the feedback since they were written in informal language (microtext)~\cite{satapathy2017phonetic} which our current version is not able to handle.
 
Due to the non-availability of sensitive customer data, Nadine cannot perform all tasks or functionalities that a customer agent can perform. We also need to add a security layer to be able to handle sensitive data. In an open social interaction, there are no restrictions on questions a customer can ask. It is challenging to train a robot for open-ended questions. Thus, if Nadine is not trained for any question, she would have to go online and depend on her network speed for her answers. This will affect her response time and performance.  It includes retraining model to answer more effectively and quick. 

The appearance of the robot plays a vital role in the way customers perceive the robot. Due to the uncanny valley, the expectation of realism from a human-like robot is high. The customers would believe Nadine could handle all types of situations and interactions like other human agents. Due to this, the results were mostly positive but with negative towards manicure and other minute details.  Few comments are related to the hardware used (such as the speaker, microphone), the language of communication and appearance (requires manicure) that can be easily changed to give a better customer interaction experience. We observe that most majority of sentiments are positive. The positive sentiments are mainly for the appearance, while negative sentiments focus around functionality, performance and response time. The negative sentiments are the result of the robot being very human-like, which increases the expectation of customers. We are also working on adding microtext normalization~\cite{satapathy2019phonsenticnet} to handle informal texts and common sense reasoning for an effective dialogue system. Future work revolves mostly around the negative aspects customers gave feedback about.

To summarize, our results show that the customers had an overall positive experience with Nadine as their service agent. Both user survey and aspect-based sentiment analysis of customer feedback show that Nadine's social behaviour was acceptable and pleasing. The functionality and performance of Nadine was limited due to some of the reasons as mentioned above but can be improved as a part of our future work.

%\subsection{Discussions}
%\subsection{Future Work}

\section{Conclusion}
We have conducted user studies on a social robot at the workplace. Based on the customer feedback and survey questionnaire, we have identified customer expectations and demands of such a robot employee. We analyzed customer feedback using aspect-based sentiment analysis to identify aspects and sentiments associated with them. From our experiments, we observed that the general customer sentiment was positive to Nadine. Functionality, appearance and performance were 3 main aspects of customer feedback. From our analysis, we also observed that Nadine performs well in a work environment and is capable of maintaining proper social etiquette that is pleasing to the customers.

%The study reveals the aspects custiomers talk about in their feedback. The overall sentiment towards a humanoid agent in workplace is positive. But, when we dig into the aspects we found the human-like fetures 
%We found that in the survey, customers found the robot to be very human-like. Their expectation in the robot were very high which led to highlight of even minor flaws like manicure.  common sense, limitation due to the environment

\section*{ACKNOWLEDGMENT}

This research is supported by the BeingTogether Centre, a collaboration between Nanyang
Technological University (NTU) Singapore and University of North Carolina (UNC) at Chapel Hill. The BeingTogether Centre is supported by the National Research Foundation, Prime Minister’s Office, Singapore under its International Research Centres in Singapore Funding Initiative. We would also like to thank our colleagues Yiep Soon and Ajay Vishwanath for their support to setup Nadine for the experiment.

%%%%%%%%%%%%%%%%%%%%%%%%%%%%%%%%%%%%%%%%%%%%%%%%%%%%%%%%%%%%%%%%%%%%%%%%%%%%%%%%

\bibliographystyle{IEEEtran}
\bibliography{ref.bib}

% Generated by IEEEtran.bst, version: 1.14 (2015/08/26)
\begin{thebibliography}{10}
\providecommand{\url}[1]{#1}
\csname url@samestyle\endcsname
\providecommand{\newblock}{\relax}
\providecommand{\bibinfo}[2]{#2}
\providecommand{\BIBentrySTDinterwordspacing}{\spaceskip=0pt\relax}
\providecommand{\BIBentryALTinterwordstretchfactor}{4}
\providecommand{\BIBentryALTinterwordspacing}{\spaceskip=\fontdimen2\font plus
\BIBentryALTinterwordstretchfactor\fontdimen3\font minus
  \fontdimen4\font\relax}
\providecommand{\BIBforeignlanguage}[2]{{%
\expandafter\ifx\csname l@#1\endcsname\relax
\typeout{** WARNING: IEEEtran.bst: No hyphenation pattern has been}%
\typeout{** loaded for the language `#1'. Using the pattern for}%
\typeout{** the default language instead.}%
\else
\language=\csname l@#1\endcsname
\fi
#2}}
\providecommand{\BIBdecl}{\relax}
\BIBdecl

\bibitem{Han_EducationRobots_HAI_2015}
\BIBentryALTinterwordspacing
J.~Han, I.-W. Park, and M.~Park, ``Outreach education utilizing humanoid type
  agent robots,'' in \emph{Proc. of the 3rd Intl. Conf. on Human-Agent
  Interaction}, ser. HAI '15.\hskip 1em plus 0.5em minus 0.4em\relax New York,
  NY, USA: ACM, 2015, pp. 221--222. [Online]. Available:
  \url{http://doi.acm.org/10.1145/2814940.2814980}
\BIBentrySTDinterwordspacing

\bibitem{Josephine_HealthRobots_SR_2015}
J.~R. Orejana, H.~S. MacDonald, Bruce A.and~Ahn, K.~Peri, and E.~Broadbent,
  ``Healthcare robots in homes of rural older adults,'' in \emph{Social
  Robotics}, A.~Tapus, E.~Andr{\'e}, J.-C. Martin, F.~Ferland, and M.~Ammi,
  Eds.\hskip 1em plus 0.5em minus 0.4em\relax Cham: Springer International
  Publishing, 2015, pp. 512--521.

\bibitem{NadineSR_CGI_2019}
M.~Ramanathan, N.~Mishra, and N.~M. Thalmann, ``{Nadine Humanoid Social
  Robotics Platform},'' in \emph{Accepted for publication in{ Computer Graphics
  International (CGI)}}.\hskip 1em plus 0.5em minus 0.4em\relax Springer, June
  2019.

\bibitem{cambria2015sentic}
E.~Cambria and A.~Hussain, ``Sentic computing,'' \emph{Cognitive Computation},
  vol.~7, no.~2, pp. 183--185, 2015.

\bibitem{Actroid}
\BIBentryALTinterwordspacing
K.~C. Ltd. (2003) Actroid. [Online]. Available:
  \url{https://en.wikipedia.org/wiki/Actroid}
\BIBentrySTDinterwordspacing

\bibitem{iCub:2011}
P.~Kormushev, S.~Calinon, R.~Saegusa, and G.~Metta, ``Learning the skill of
  archery by a humanoid robot icub,'' in \emph{Proc. IEEE intl conf. on
  humanoid robots (humanoids), Nashville}, 01 2011, pp. 417 -- 423.

\bibitem{Giuliani:2013}
\BIBentryALTinterwordspacing
M.~Giuliani, R.~P. Petrick, M.~E. Foster, A.~Gaschler, A.~Isard, M.~Pateraki,
  and M.~Sigalas, ``Comparing task-based and socially intelligent behaviour in
  a robot bartender,'' in \emph{Proceedings of the 15th ACM on International
  Conference on Multimodal Interaction}, ser. ICMI '13.\hskip 1em plus 0.5em
  minus 0.4em\relax New York, NY, USA: ACM, 2013, pp. 263--270. [Online].
  Available: \url{http://doi.acm.org/10.1145/2522848.2522869}
\BIBentrySTDinterwordspacing

\bibitem{museum_guide:2005}
M.~Bennewitz, F.~Faber, D.~Joho, M.~Schreiber, and S.~Behnke, ``Towards a
  humanoid museum guide robot that interacts with multiple persons,'' in
  \emph{5th IEEE-RAS International Conference on Humanoid Robots, 2005.}\hskip
  1em plus 0.5em minus 0.4em\relax IEEE, 2005, pp. 418--423.

\bibitem{Tutors_robot:2004}
T.~Kanda, T.~Hirano, D.~Eaton, and H.~Ishiguro, ``Interactive robots as social
  partners and peer tutors for children: A field trial,'' \emph{Human Computer
  Interaction (Special issues on human-robot interaction)}, vol.~19, pp.
  61--84, 06 2004.

\bibitem{Ljungblad:2012}
\BIBentryALTinterwordspacing
S.~Ljungblad, J.~Kotrbova, M.~Jacobsson, H.~Cramer, and K.~Niechwiadowicz,
  ``Hospital robot at work: Something alien or an intelligent colleague?'' in
  \emph{Proceedings of the ACM 2012 Conference on Computer Supported
  Cooperative Work}, ser. CSCW '12.\hskip 1em plus 0.5em minus 0.4em\relax New
  York, NY, USA: ACM, 2012, pp. 177--186. [Online]. Available:
  \url{http://doi.acm.org/10.1145/2145204.2145233}
\BIBentrySTDinterwordspacing

\bibitem{Robotics_Waiter}
A.~Cheong, E.~Foo, M.~Lau, J.~Chen, and H.~Gan, ``Development of a robotics
  waiter system for the food and beverage industry,'' in \emph{3rd
  International Conference On Advances in Mechanical \& Robotics
  Engineering}.\hskip 1em plus 0.5em minus 0.4em\relax Newcastle University,
  2015, pp. 21--25.

\bibitem{Robonaut}
M.~A. Diftler, J.~Mehling, M.~E. Abdallah, N.~A. Radford, L.~B. Bridgwater,
  A.~M. Sanders, R.~S. Askew, D.~M. Linn, J.~D. Yamokoski, F.~Permenter
  \emph{et~al.}, ``Robonaut 2-the first humanoid robot in space,'' in
  \emph{2011 IEEE international conference on robotics and automation}.\hskip
  1em plus 0.5em minus 0.4em\relax IEEE, 2011, pp. 2178--2183.

\bibitem{MOIRA}
K.~Osuka and H.~Kitajima, ``Development of mobile inspection robot for rescue
  activities: Moira,'' in \emph{Proceedings 2003 IEEE/RSJ International
  Conference on Intelligent Robots and Systems (IROS 2003)(Cat. No.
  03CH37453)}, vol.~4.\hskip 1em plus 0.5em minus 0.4em\relax IEEE, 2003, pp.
  3373--3377.

\bibitem{Robot_Motivate_Exercise_Older}
J.~{Fasola} and M.~J. {Mataric}, ``Using socially assistive human–robot
  interaction to motivate physical exercise for older adults,''
  \emph{Proceedings of the IEEE}, vol. 100, no.~8, pp. 2512--2526, Aug 2012.

\bibitem{robot-waiter}
\BIBentryALTinterwordspacing
(2016) Waiter robot. [Online]. Available:
  \url{https://www.thejournal.ie/robot-waiter-2709346-Apr2016/}
\BIBentrySTDinterwordspacing

\bibitem{Mining_summarizing_customer_reviews}
\BIBentryALTinterwordspacing
M.~Hu and B.~Liu, ``Mining and summarizing customer reviews,'' in
  \emph{Proceedings of the Tenth ACM SIGKDD International Conference on
  Knowledge Discovery and Data Mining}, ser. KDD '04.\hskip 1em plus 0.5em
  minus 0.4em\relax New York, NY, USA: ACM, 2004, pp. 168--177. [Online].
  Available: \url{http://doi.acm.org/10.1145/1014052.1014073}
\BIBentrySTDinterwordspacing

\bibitem{Product_feature_Scoring_from_Reviews}
\BIBentryALTinterwordspacing
C.~Scaffidi, K.~Bierhoff, E.~Chang, M.~Felker, H.~Ng, and C.~Jin, ``Red opal:
  Product-feature scoring from reviews,'' in \emph{Proceedings of the 8th ACM
  Conference on Electronic Commerce}, ser. EC '07.\hskip 1em plus 0.5em minus
  0.4em\relax New York, NY, USA: ACM, 2007, pp. 182--191. [Online]. Available:
  \url{http://doi.acm.org/10.1145/1250910.1250938}
\BIBentrySTDinterwordspacing

\bibitem{Weakness_Finder}
W.~Zhang, H.~Xu, and W.~Wan, ``Weakness finder: Find product weakness from
  chinese reviews by using aspects based sentiment analysis,'' \emph{Expert
  Systems with Applications}, vol.~39, pp. 10\,283 -- 10\,291, 09 2012.

\bibitem{Chatterbot_2018}
\BIBentryALTinterwordspacing
Gunthercox. (2018) Chatterbot. [Online]. Available:
  \url{https://chatterbot.readthedocs.io/en/stable/}
\BIBentrySTDinterwordspacing

\bibitem{hukdd}
\BIBentryALTinterwordspacing
M.~Hu and B.~Liu, ``Mining and summarizing customer reviews,'' in
  \emph{Proceedings of the Tenth ACM SIGKDD International Conference on
  Knowledge Discovery and Data Mining}, ser. KDD '04.\hskip 1em plus 0.5em
  minus 0.4em\relax New York, NY, USA: ACM, 2004, pp. 168--177. [Online].
  Available: \url{http://doi.acm.org/10.1145/1014052.1014073}
\BIBentrySTDinterwordspacing

\bibitem{poria2016aspect}
S.~Poria, E.~Cambria, and A.~Gelbukh, ``Aspect extraction for opinion mining
  with a deep convolutional neural network,'' \emph{Knowledge-Based Systems},
  vol. 108, pp. 42--49, 2016.

\bibitem{cambria2018senticnet}
E.~Cambria, S.~Poria, D.~Hazarika, and K.~Kwok, ``Senticnet 5: Discovering
  conceptual primitives for sentiment analysis by means of context
  embeddings,'' in \emph{Thirty-Second AAAI Conference on Artificial
  Intelligence}, 2018, pp. 1795 -- 1802.

\bibitem{vilares2018babelsenticnet}
D.~Vilares, H.~Peng, R.~Satapathy, and E.~Cambria, ``Babelsenticnet: a
  commonsense reasoning framework for multilingual sentiment analysis,'' in
  \emph{2018 IEEE Symposium Series on Computational Intelligence (SSCI)}.\hskip
  1em plus 0.5em minus 0.4em\relax IEEE, 2018, pp. 1292--1298.

\bibitem{satapathy2017phonetic}
R.~Satapathy, C.~Guerreiro, I.~Chaturvedi, and E.~Cambria, ``Phonetic-based
  microtext normalization for twitter sentiment analysis,'' in \emph{2017 IEEE
  International Conference on Data Mining Workshops (ICDMW)}.\hskip 1em plus
  0.5em minus 0.4em\relax IEEE, 2017, pp. 407--413.

\bibitem{satapathy2019phonsenticnet}
R.~Satapathy, A.~Singh, and E.~Cambria, ``Phonsenticnet: A cognitive approach
  to microtext normalization for concept-level sentiment analysis,''
  \emph{arXiv preprint arXiv:1905.01967}, 2019.

\end{thebibliography}

\end{document}